# On applications of conservation laws in pharmacokinetics

S.Piekarski (1), M.Rewekant (2)

Institute of Fundamental Technological Research Polish Academy of Sciences (1), Medical University of Warsaw, Poland

**Abstract**

There has been certain criticism raised by A. Rescigno [2,8,9,12-15] against the standard formulation of pharmacokinetics.

In 2011 it has been suggested that inconsistencies in pharmacokinetics should be eliminated after deriving "pharmacokinetic parameters" from conservation laws [3]. In the following text a simple system of conservation laws for extra - vascular administration of a drug is explicitly given and preliminary discussion concerning this issue is included.

**Introduction**

Some criticism against the standard formulation of pharmacokinetics has been raised by A.Rescigno [2,8,9,12-15].

In 2011 it has been suggested that inconsistencies should be eliminated in pharmacokinetics after deriving "pharmacokinetic parameters" from conservation laws [3]. A simple example of a system of conservation laws is given in [4].

However, Piekarski et al. describe only intravenous administration of a drug while the problem of extravenous administration is more interesting [4].

Therefore, in the following text the equations for the extravascular administration are explicitly given.

In principle, in order to determine parameters in our model one should apply the nonlinear regression and we hope to do it in the future.

It should be mentioned that our equations are nonlinear because of the effects of protein – drug binding and, in general, describe processes that are outside temporal chemical equilibrium.

The important feature of our equations is that both unbound plasma concentration of a drug and bound plasma concentration of a drug are treated separately.

In principle, the values of parameters in our equations should be determined by the nonlinear regression and we hope to do that in the future.

## Basic Equations

The standard equations of chemical kinetics relating to free drug concentration $s(t)$, protein concentration $e(t)$ and the drug – protein complex $c(t)$ are as follows:

$$\frac{\partial s(t)}{\partial t} = -k_+ e(t)s(t) + k_- c(t) \qquad (1)$$

$$\frac{\partial e(t)}{\partial t} = -k_+ e(t)s(t) + k_- c(t) \qquad (2)$$

$$\frac{\partial c(t)}{\partial t} = k_+ e(t)s(t) - k_- c(t) \qquad (3)$$

where all quantities depend on time $t$ and $k_+$, $k_-$ are on and off rate. ($k_+ > 0$, $k_- > 0$).

The above equations describe the time evolution in a uniform system without external sources.

Now, the elimination of a drug is modelled - according to the compartmental pharmacokinetics - by the additional source term in the balance law (1). In the simplest possible version, such additional source term can by modelled by the term proportional to $s(t)$ and the "strength" of the elimination process can be measured by the proportionality coefficient $\alpha$. Therefore, the resulting equations are now:

$$\frac{\partial s(t)}{\partial t} = -k_+ e(t)s(t) + k_- c(t) - \alpha s(t) \qquad (4)$$

$$\frac{\partial e(t)}{\partial t} = -k_+ e(t)s(t) + k_- c(t) \qquad (5)$$

$$\frac{\partial c(t)}{\partial t} = k_+ e(t)s(t) - k_- c(t) \qquad (6)$$

In order to describe the extravascular administration one has to add the additional term describing the rate of absorption of a drug into the plasma (compare [2]). This term shall be denoted

$$Q(t)$$

and the whole system (4) – (6) takes the modified form:

$$\frac{\partial s(t)}{\partial t} = -k_+ e(t)s(t) + k_- c(t) - \alpha s(t) + Q(t) \tag{7}$$

$$\frac{\partial e(t)}{\partial t} = -k_+ e(t)s(t) + k_- c(t) \tag{8}$$

$$\frac{\partial c(t)}{\partial t} = k_+ e(t)s(t) - k_- c(t) \tag{9}$$

The equations (7) – (9) still do not take into account the transport processes between the central compartment and the peripheral compartment.

Let $V_0$ denote the volume of the central compartment (usually between 5 or 6 litres): then the number of molecules of a free fraction of a drug in the central compartment is $V_0 s(t)$.

Let $N(t)$ denote the number of molecules of a drug in the peripheral compartment.

In order to formulate the simplest possible dynamical model of the transport processes between both compartments let us assume that the elimination of a drug is „switched out".

The resulting conservation law for the transport of the free fraction of a drug between both compartments is:

$$\frac{\partial}{\partial t}[V_0 s(t) + N(t)] = 0 \tag{10}$$

Every model of the transport processes between both compartments should identically satisfy the condition (10). The simplest model depends on two real parameters and assumes that the volume of the peripheral compartment is $V_{eff}$ and that $N(t)$ molecules are distributed with a uniform density $n(t)$

$$N(t) = V_{eff} n(t) \tag{11}$$

The explicit form of this model is

$$\frac{\partial}{\partial t} V_0 s(t) = \beta [n(t) - s(t)] \tag{12}$$

$$\frac{\partial}{\partial t} V_{eff} n(t) = -\beta [n(t) - s(t)] \tag{13}$$

where the real parameter $\beta$ models the rate of the transport processes between both compartments.

Now one can see that the final system of evolution equations for the drug transport is:

$$\frac{\partial s(t)}{\partial t} = -k_+ e(t) s(t) + k_- c(t) - \alpha s(t) + \frac{\beta}{V_0} [n(t) - s(t)] + Q(t) \tag{14}$$

$$\frac{\partial e(t)}{\partial t} = -k_+ e(t) s(t) + k_- c(t) \tag{15}$$

$$\frac{\partial c(t)}{\partial t} = k_+ e(t) s(t) - k_- c(t) \tag{16}$$

$$\frac{\partial}{\partial t} V_{eff} n(t) = -\beta [n(t) - s(t)] \tag{17}$$

In the future we hope to interpret the solutions of this system in terms of experimental results.

Obviously, in general similar models can contain many different compartments [1-16].

## Conclusions

It is important to compare intravenous and extravascular administration of a drug by means of nonlinear regression.

In principle, our approach (based on the conservation laws) can be applied to modelling of bioavailability and bioequivalence. Certainly, before that one should model AUCs corresponding to our systems of conservation laws.

We hope to continue such discussions in the future.